\documentclass[twocolumn,aps,prl]{revtex4}

\usepackage{amsmath,amsfonts,amssymb}
\usepackage{wrapfig}
\usepackage{graphicx}
\usepackage{bbm}
\usepackage{color}
\usepackage{subfigure}

\usepackage[english]{babel}
\makeatletter

\def\compoundrel#1\over#2{\mathpalette\compoundreL{{#1}\over{#2}}}
\def\compoundreL#1#2{\compoundREL#1#2}
\def\compoundREL#1#2\over#3{\mathrel
      {\vcenter{\hbox{$\m@th\buildrel{#1#2}\over{#1#3}$}}}}
\makeatother

\usepackage{babel}
\makeatother

\def\be{\begin{eqnarray}}
\def\ee{\end{eqnarray}}
\def\bee{\begin{eqnarray*}}
\def\eee{\end{eqnarray*}}

\begin{document}

\title{Intra-molecular refrigeration in enzymes}

\author{Hans J. Briegel$^{1,2}$ and Sandu Popescu$^{3,4}$}

\affiliation{$^1$Institut f{\"u}r Theoretische Physik,
Universit{\"a}t Innsbruck, Technikerstra{\ss }e 25, A-6020 Innsbruck\\
$^2$ Institut f{\"u}r Quantenoptik und Quanteninformation der
\"Osterreichischen Akademie der Wissenschaften, Innsbruck, Austria\\
$^3$ H.H. Wills Physics Laboratory, University of Bristol, Tyndall
Avenue, Bristol BS8 1TL, U.K.\\
$^4$ Hewlett-Packard Laboratories, Stoke Gifford, Bristol BS12 6QZ,
U.K.}

\begin{abstract}
We present a simple mechanism for intra-molecular refrigeration, where parts of a molecule are actively cooled below the environmental temperature. We discuss the potential role and applications of such a mechanism in biology, in particular in enzymatic reactions.
\end{abstract}

\maketitle

{\bf Introduction}. In the present paper, we would like to suggest the possibility that enzymes can cool their active sites below the environmental temperature and thereby increase their functionality. A similar cooling mechanism might be used by molecular machines to cool parts of the machine and to reduce the deteriorating effects of thermal noise. The very concept of intra-molecular refrigeration was proposed by us in a recent paper \cite{BriegelPopescu0806}. Here, we discuss it in more detail and present, for the first time, a possible molecular mechanism.

Let us first discuss why be believe that molecular cooling is important and what functionality it could have in biological systems. To start with, we note that for many species, including mammals and birds, the ability to cool parts of their body is essential to survive. Their body temperature is kept largely stable at a specific temperature even when they live in an environment with varying temperatures. This is only possible since they have a built-in cooling system. Such a cooling mechanism works however at a large scale - essentially at the scale of the whole multicellular organism and generally involves the interaction of many organs. The type of cooling we are interested here is, however, completely different and could, in principle, exist in all three domains of life - bacteria, archaea and eukarya.  We suggest the possibility that cooling below environmental temperature may occur at molecular level.

\begin{figure}[htb]
\begin{center}
\begin{minipage}{9cm}
\hspace{-0.1cm}
\subfiguretopcaptrue
\subfigure[]{
\includegraphics[width=4cm]{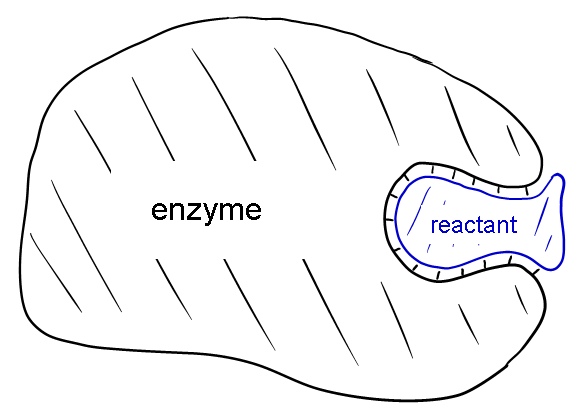}}
\hspace{0.5cm}
\subfiguretopcaptrue
\subfigure[]{
\includegraphics[width=4cm]{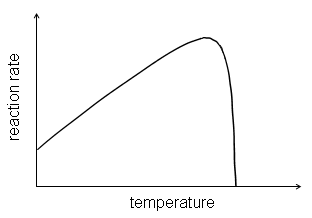}}
\end{minipage}
\end{center}
\caption{(a) Molecular model of catalysis: Binding of a reactant molecule (blue) to the active site of the enzyme. (b) Temperature dependence of a chemical reaction which is catalyzed by an enzyme.} \label{EnzymeModel}
\end{figure}

If cooling could be achieved at molecular level, this would have obvious benefits. An example is an increased efficiency of catalysis: Many proteins act as catalyzers (enzymes) with very high specificity \cite{Alberts08}. They have active sites in the shape of cavities, which bind only molecules that fit precisely into the cavity, like a hand in a glove [see Fig.~\ref{EnzymeModel}(a)]. When a reactant molecule (substrate) binds to the active site, its activation energy to react with some other molecule is lowered, and the reaction is thus sped-up due to the presence of the enzyme. After the reaction has taken place, the reactant molecule (product) leaves the cavity. The temperature dependence of the reaction rate in the presence of an enzyme typically looks like the curve displayed in Fig.~\ref{EnzymeModel}(b). At moderate temperatures, the reaction rate increases with the temperature through an increased thermal energy of the reactant molecules.
At higher temperature, however, vibrations in the protein may lead to a deformation of the cavity, which leads in turn to a reduced efficiency to bind reactant molecules (and eventually to de-naturation) and hence to a decreased rate of reaction.
From this description \cite{Alberts08} it is clear that, if an enzyme would succeed in cooling its active site below the environmental temperature by no-matter-what mechanism, then the enzymatic reaction will become more resistant and functionality will be increased. Cooling could, in principle, be achieved by different methods. The model we describe here is based on conformational changes, which is a common feature in molecular dynamics.

\vspace*{3pt}{\bf Molecular mechanism}.
Let us now describe the molecular model in a little more detail. As we have mentioned, the basic mechanism that is responsible for catalysis is the binding of the substrate to the active site of the enzyme, which has the effect of lowering the activation energy for the chemical reaction.

\begin{figure}[htb]
\begin{minipage}{7cm}
\hspace{-0.1cm}
(a)\\
\includegraphics[width=6cm]{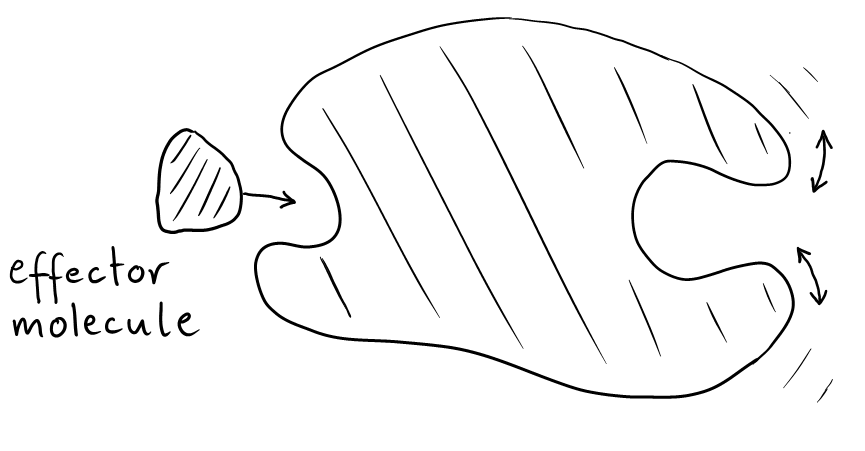}\\
\vspace{0.25cm}
(b)\\
\hspace{1.8cm}
\includegraphics[width=4.8cm]{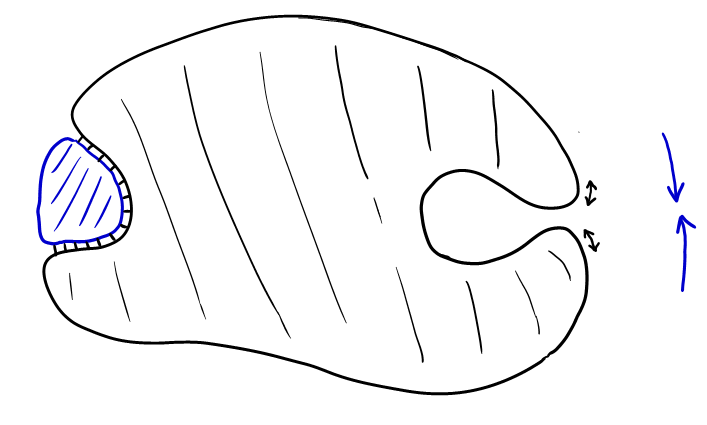}\\
\vspace{0.25cm}
(c)\\
\hspace{0.3cm}
\includegraphics[width=5.2cm]{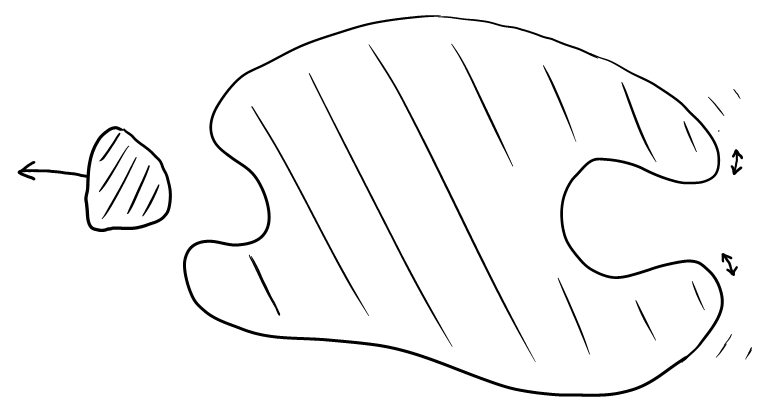}
\end{minipage}
\caption{Illustration of refrigeration via allosteric conformational change.
(a) At very high temperatures, deformation and vibrations of the cavity at the active site will prevent reactant molecules to bind to the enzyme.
(b) Cooling of the active site can be achieved through a conformational change that contracts the cavity temporarily. Since in the ``closed'' position the cavity is stiffer, at thermal equilibrium the amplitude of vibrations is smaller than in the ``open'' position.  Free energy needs to be supplied, e.g. by ATP or a similar effector molecule, docking to another active (so-called allosteric) site.
(c) Upon reopening of the cavity, the amplitude of vibration remains reduced, until re-thermalisation.}
\label{RefrigerationProcess}
\end{figure}

The shape of the cavity at the active site is defined by a specific arrangement of functional groups, which matches with the molecular structure of the substrate. When the ambient temperature increases, there will be more frequent (and higher energetic) collisions of reactant molecules with the enzyme which increases the reaction rate. However, when the temperature becomes too high, it will lead to increased thermal vibrations of the functional groups near the active site as indicated in Fig.~\ref{RefrigerationProcess}(a). As a consequence, the substrate will no longer be able to bind to the enzyme and the reaction stops.

The simple refrigeration mechanism we propose is based on a conformational change of the enzyme, which contracts the cavity at the active site and thereby stabilizes its vibrating parts. The reason is that, if the cavity contracts, its walls will interact with each other, making the whole cavity ``stiffer''. Suppose the cavity stays closed until it reaches thermal equilibrium. At equilibrium, due to their mutual interaction, the amplitude of the movement of the walls relative to each other is reduced as compared to what it was when the cavity was open, see Fig.~\ref{RefrigerationProcess}(b).
Indeed, we can see this in a different way: if the amplitude of the oscillation in the stiffer configuration would be the same as in the open configuration, it would cost more energy, i.e.\ above the average energy corresponding to the ambient temperature. Suppose now the cavity re-opens at a speed fast relative to the thermalization rate, but slow compared to the period of vibration. In this case, the opening does not itself amplify the amplitude of vibration. Hence the cavity ends up in the open position with reduced vibrations, i.e.\ at a lower temperature, as is indicated in Fig.~\ref{RefrigerationProcess}(c).
The cavity is now cold and ready to accept new reactants.

As in any refrigerator, in order for cooling to work, there must be a supply of free energy. We suggest as a possible mechanism that the conformational change is induced by an allosteric process, in which the effector molecule supplies the free energy, as illustrated in Fig.~\ref{RefrigerationProcess}. However, we emphasize that the essential part of the proposal is the refrigeration via conformational change. How exactly the conformational change is produced, and how the free energy is delivered, is irrelevant.

\begin{figure}[htb]
\begin{center}
\begin{minipage}{9cm}
\hspace{-0.5cm}
\subfiguretopcaptrue
\subfigure[]{
\includegraphics[width=4cm]{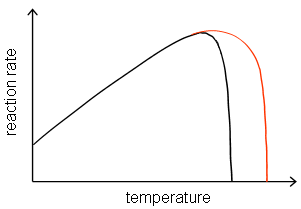}}
\hspace{0.5cm}
\subfigure[]{
\includegraphics[width=4cm]{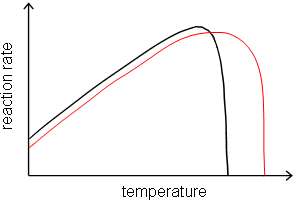}}
\end{minipage}
\end{center}
\caption{Proposal of an experiment. (a) Predicted temperature dependence of the rate of an enzymatic reaction, without and with (red) supply of effector molecules. (b) For lower temperatures, the cooling may have a detrimental effect on the enzyme efficiency.}
\label{ExperimentalProposal}
\end{figure}

\vspace*{3pt}{\bf Experiment}. An experimental demonstration of molecular cooling could be straightforward. Suppose one finds a catalyzing enzyme with an active site as described. Several measurement rounds are then performed, observing the enzyme activity (reaction rate) as a function of temperature. In each round, the same concentrations of enzyme and substrate are used, but the concentration of the effector molecules is varied. The predicted temperature dependence of the chemical reaction rate would then be as described in Fig.~\ref{ExperimentalProposal}(a)(red curve). The enzyme will retain its function up to higher temperatures due to the molecular cooling effect when effector molecules are present, compared to the situation when there are no effectors. It is also possible that for lower temperatures the cooling may have a detrimental effect on the enzyme efficiency, since during the time when the active site is closed, it cannot accept any substrate, see Fig.~\ref{ExperimentalProposal}(b).

\vspace*{3pt}{\bf Physical model}. Although we talk about cooling at a molecular level, the refrigeration mechanism is essentially classical, that is, not quantum mechanical. For our purposes, the different parts of the enzyme constituting the active site can be viewed as mechanical systems of different masses, which can oscillate around their equilibrium configuration. At increasing temperatures, the different parts of the system start vibrating with increasing amplitudes, which diminishes the functionality of the enzyme.

The active site of the enzyme can vibrate in many complicated ways. Technically, there are many vibration modes. For simplicity, here we analyze the case of just two oscillators, which stresses the essential features of the model. The two oscillating parts could represent the two ``lips'' of the cavity in Fig.~\ref{RefrigerationProcess}(a). The vibrations indicated in the figure then correspond to excitations of the oscillators around their respective equilibrium positions.
Contracting the cavity brings the parts together; when the two lips are closer, additional forces will act on them. Whatever the sign of these forces (attractive or repulsive) will be, the net effect is to make the entire system tighter and increase the frequency of the vibrations.

This process can be modeled by an extra interaction potential which is switched on in the contracted configuration. While the exact nature of these interactions is not essential, its main effect is to increase the frequency of the oscillators which will be energetically more difficult to excite.

The Hamilton function for this system is given by
\begin{equation}
H(t)=\frac{P_1^2}{2m_1}+\frac{P_2^2}{2m_2}+\frac{\kappa_1}{2}X_1^2
+\frac{\kappa_2}{2}X_2^2+\frac{\kappa(t)}{2}\left(X_2-X_1\right)^2
\label{Hamiltonian1}
\end{equation}
where $X_i$ is the displacement of mass $m_i$ from its equilibrium position, $P_i$ the corresponding canonical momentum, and $\kappa_i$ the spring constant associated with the $i$th oscillator. The last term describes the variable interaction potential, which is switched on in the contracted configuration, accounted for by the variable spring constant $\kappa(t)$. Since we are only interested in the system near the equilibrium, we can choose quadratic oscillator potentials for all the involved potentials.  For simplicity, we shall assume equal masses and spring constants for both oscillators, i.e. $m_1=m_2=m$ and $\kappa_1=\kappa_2=\kappa$. It is convenient to introduce new variables, $X_{\rm cm}=(X_1+X_2)/2$, $P_{\rm cm}=P_1+P_2$  for the center of mass, and $x=X_1-X_2$ , $p=(P_2-P_1)/2$  for the relative displacement coordinates and momenta. The transformed Hamilton function then reads
\begin{equation}
H(t)=\frac{P_{\rm cm}^2}{2M}+\frac{\kappa_{\rm cm}}{2}X_{\rm cm}^2
+\frac{p^2}{2\mu}+\frac{\kappa_{\rm rel}+\kappa(t)}{2}x^2
\label{Hamiltonian2}
\end{equation}
with $\kappa_{\rm cm}=2\kappa$, $\kappa_{\rm rel}=\kappa/2$, $M=2m$, $\mu=m/2$. The motions of the centre of mass and the relative displacement coordinates are decoupled and describe harmonic oscillations. While the first does not lead to a deformation of the cavity, the latter does. We will thus concentrate on the motion of the relative displacements in the following, described by
\begin{equation}
H(t)=\frac{p^2}{2\mu}+\frac{\kappa_{\rm rel}+\kappa(t)}{2}x^2 \,.
\label{Hamiltonian3}
\end{equation}
When the cavity is open, $\kappa(t)\simeq 0$, and, due the influence of the environment, the relative mode of oscillation will be thermally excited. Upon contracting the cavity, the coupling function $\kappa(t)$ will effectively be switched on and tighten the potential. This corresponds to an increase of the spring constant, which makes this oscillator stiffer. To understand how this can lead to cooling, it is convenient to use a quantum mechanical description of the oscillator, where the energies are quantized. We would like to emphasize however that we consider a situation that is essentially classical, that is, we are far away from the quantum mechanical ground state. Using the quantum mechanical language is here simply a technical convenience; we could also have solved the system classically.

The Hamilton operator corresponding to (\ref{Hamiltonian3}) can be written in the form
\begin{equation}
H(t)=\hbar\omega(t)\left( a^{\dag}a + \frac{1}{2}\right)
\label{Hamiltonian4}
\end{equation}
with a time-dependent frequency
\begin{equation}
\omega(t)=\sqrt{[\kappa_{\rm rel}+\kappa(t)]/\mu},
\end{equation}
and ladder operators
\begin{eqnarray}
a &=& a_t=\frac{x}{\sqrt{2\hbar/\mu\omega(t)}}
+i\frac{p}{\sqrt{2\hbar\omega(t)\mu}} \\
a^{\dag}&=& a_t^{\dag}=\frac{x}{\sqrt{2\hbar/\mu\omega(t)}}
-i\frac{p}{\sqrt{2\hbar\omega(t)\mu}}.
\end{eqnarray}
For a fixed frequency, $\omega(t)=\omega={\rm const}$, the possible energies of this oscillator are given by the formula $E_n=\hbar\omega(n+1/2)$, $n=1,2,...$. In a thermal environment at temperature $T$, the different energies $E_n$ will be found with probabilities $p_n=Z^{-1}\exp(-n\hbar\omega/k_BT)$, with $k_B$ and $\hbar$ denoting the Boltzmann and Planck constant, resp., and $Z$ a normalization factor (called the partition function) which ensures $\sum_n p_n=1$.

\begin{figure}[htb]
\begin{center}
\includegraphics[width=8cm]{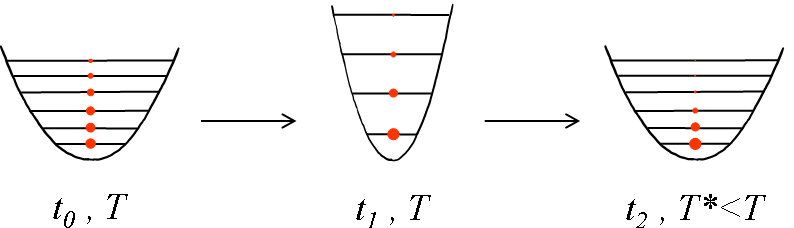}
\end{center}
\caption{Refrigeration of the molecular vibrations through a conformational change (see text).}
\label{PotentialVariation}
\end{figure}

In the open and contracted configuration, the oscillator will have two different frequencies, $\omega_0$ and $\omega_1 > \omega_0$, respectively. Suppose that the system starts at time $t_0$  in the open configuration, with a spacing of the allowed energies $\Delta E_n^{0}=\hbar\omega_0$, and with a thermal-state distribution
$p_n(t_0)=Z^{-1}\exp(-n\hbar\omega_0/k_BT)$. Upon contraction, the spacing between the allowed energies will increase to $\Delta E_n^{1}=\hbar\omega_1 > \hbar\omega_0$ (see Fig.~\ref{PotentialVariation}). The thermalization of the oscillator is a process with a characteristic time scale that depends on the interaction strength between the system and the environment. Ideally we assume that, after contraction, the system stays long enough in this configuration, such that it has time to thermalize. The resulting thermal distribution at time $t_1$ is then
$p_n(t_1)=Z^{-1}\exp(-n\hbar\omega_1/k_BT)$.
Note that, since the spacing between the energy levels has increased, a larger fraction of the population will move to lower energies, as indicated in Fig.~\ref{PotentialVariation}.

Upon re-opening, the system will move back to the open configuration (with frequency $\omega_0$). Let the time required for opening be $\tau_{\rm open}$. Whether or not cooling takes place depends on two time scales. The first is the thermalization time   $\tau_{\rm therm}$ and the second the period of oscillation $\tau_{\rm osc}=2\pi\omega^{-1}$. The ideal regime is when the transition is fast compared to thermalization but slow compared to the oscillation period, i.e.\ $\tau_{\rm osc}\ll \tau_{\rm open}\ll \tau_{\rm therm}$. In that case the system will cool down. Indeed, in this regime the opening is adiabatic and the relative population of the different energy levels does not change during the transition. In this ideal case, right after returning to the open configuration, i.e.\ at time $t_2=t_1+\tau_{\rm open}$, the population of the energy levels will still be the same as at time $t_1$, that is $p_n(t_2)=Z^{-1}\exp(-n\hbar\omega_1/k_BT)$ $=$ $Z^{-1}\exp(-n\hbar\omega_0/k_BT^{\star})$. With respect to the new energy spectrum (spacing $\Delta E_n^{0}=\hbar\omega_0$), this corresponds however to a thermal state with reduced temperature $T^{\star}=\frac{\omega_0}{\omega_1}T$. After such a cooling cycle, the temperature of the active site of the enzyme has thus been reduced by the amount $\Delta T/T = 1-\omega_0/\omega_1$. We note however that, as shown later, the cooling effect takes place, albeit with reduced efficiency, also outside the ideal regime described above.

In the remainder of this paper, we illustrate the essential dynamics of the refrigeration process for a time-dependent oscillator frequency $\omega(t)$. For the present purpose, it suffices to describe the thermalizing effect of the environment by a master equation of the form
\begin{equation}
\frac{\partial}{\partial t}\rho(t) = -\frac{i}{\hbar}\left[ H(t),\rho(t) \right] + L_t\rho(t)
\end{equation}
with $H(t)$ as in (\ref{Hamiltonian4}) and with a dissipative term
\begin{eqnarray}
L_t\rho(t) &=&
-\frac{\gamma}{2}(\nu+1)\left\{a^{\dag}a\rho+\rho a^{\dag}a - 2a\rho a^{\dag}\right\} \nonumber \\
& &
-\frac{\gamma}{2}\nu\left\{aa^{\dag}\rho+\rho aa^{\dag} - 2a^{\dag}\rho a\right\}\,.
\end{eqnarray}
Here $\rho(t)$ is the state (density matrix) of the system at time $t$. The parameter  $\gamma$ is the thermal relaxation rate that describes how fast the system would approach, for any fixed configuration ($\omega = {\rm const}$), its thermal equilibrium state
\begin{equation}
\rho_{\rm therm}(\omega)=\rho(t\to\infty)\vert_{\omega={\rm const}}=Z^{-1}\exp\left(
-\frac{\hbar\omega}{k_BT}a^{\dag}a \right)\, ,
\end{equation}
and $\nu $ is the mean number of thermal excitations in that equilibrium state,
$\nu=\nu(\omega)=1/[\exp(\hbar\omega/k_BT)-1]$.
For a time-dependent oscillator frequency, $\omega=\omega(t)$, this ``instantaneous'' thermal state thus changes with time, as does $\nu=\nu[\omega(t)]$ . Unless the motion of the oscillator is very slow compared to the relaxation time, it will be driven out of equilibrium, which gives rise to the described refrigeration effect. In the regime where the oscillation period is short compared to all other timescales  (adiabatic regime), $\tau_{\rm osc}\ll \tau_{\rm open}, \tau_{\rm therm}$, the solution can be written in the form \cite{Englert94}
\begin{equation}
\rho(t)=\frac{1}{\eta(t)}\left[ 1-\frac{1}{\eta(t)}\right]^{a^{\dag}a}
\label{QuenchedBoltzmannState}
\end{equation}
where $\eta(t)$ is
a solution of the differential equation
\begin{equation}
\frac{{\rm d}}{{\rm d}t}\eta(t) = -\gamma\eta(t) + \gamma \left\{\nu[\omega(t)]+1 \right\}
\end{equation}
which can be readily integrated to give
\begin{equation}
\eta(t) = e^{-\gamma(t-t_1)}\eta(t_1)
+ \gamma \int_{t_1}^{t}{\rm d}s\, \frac{e^{-\gamma(t-s)}}{1-e^{\frac{-\hbar\omega(s)}{k_BT}}}\,.
\label{Eta}
\end{equation}
The function $\eta(t)$ has a simple interpretation, it is equal to the mean number of excitations (plus unity) in the oscillator at time $t$, i.e.\
$\eta(t)=\langle a^{\dag}a \rangle + 1$. The state (\ref{QuenchedBoltzmannState}) describes a ``quenched'' Boltzmann distribution, corresponding to a quasi-thermal state, albeit with a temperature $T(t)$ reduced below the temperature $T$ of the environment,
\begin{equation}
\frac{T(t)}{T} = \frac{\log\left[ \nu(t)/(\nu(t)+1) \right]}{\log\left[ 1-1/\eta(t) \right]} = \frac{\log\left[ \nu(t)/(\nu(t)+1) \right]}{\log\left[ \langle a^{\dag}a \rangle /(\langle a^{\dag}a \rangle + 1) \right]} \,.
\end{equation}

\begin{figure}[t]
\begin{center}
\hspace*{-0.5cm}
\begin{minipage}{9cm}
(a)\\
\rotatebox[origin=c]{90}{\hspace*{4cm}$\frac{\omega(t)}{\omega_{1}}$, $\frac{T(t)}{T}$}\hspace{0.25cm}
\includegraphics[width=6cm]{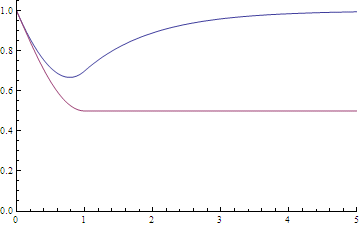}\\
\vspace*{-2.3cm}$t/\tau_{\rm open}$\\
\vspace{1cm}
(b)\\
\rotatebox[origin=c]{90}{\hspace*{4cm}$\langle a^{\dag}a \rangle$}
\hspace{0.25cm}\includegraphics[width=6cm]{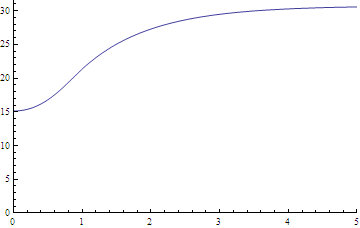}\\
\vspace*{-2cm}$t/\tau_{\rm open}$
\end{minipage}
\end{center}
\caption{Refrigeration of the active site as a function of time. (a) Temperature $T(t)/T$ of the active site (blue) and frequency of vibrations $\omega(t)/\omega_{1}$ (red). (b) Average number of quanta $\langle a^{\dag}a \rangle(t)$ in the vibration mode. [Parameters: $\omega_1/\omega_0=2$, $\gamma\tau_{\rm open}=1$, $\frac{\hbar\omega_0}{k_{B}T}=0.032$. At $T=300K$, this corresponds to  $\tau_{\rm osc} = 2\pi/\omega_{0}=5$ps as the time scale of elastic vibrations.]}
\label{RefrigerationDynamicsC}
\end{figure}

Let us now have a closer look at the essential part of the process, that is, when the molecule moves from the closed configuration (after thermalization) at time $t_1$  back to the open configuration, which it reaches at time $t_2=t_1+\tau_{\rm open}$. In Fig.~\ref{RefrigerationDynamicsC}, we plot the temperature of the oscillator - representing the active site - as a function of time.  As an example, we choose the specific time profile $\omega(t)=\omega_1+(\omega_0-\omega_1)(\sin(\frac{\pi}{2}\frac{t}{\tau_{\rm open}}))$,
for $0 = t_1\le t \le t_2 = \tau_{\rm open}$ and $\omega(t)=\omega_0$ for $t > \tau_{\rm open}$,
but other profiles give similar results. The function $\eta(t)$ in (\ref{Eta}), like $\omega(t)$, can be expressed in terms of the dimensionless quantities $\frac{\omega_1}{\omega_0}$, $\frac{\hbar\omega_0}{k_{B}T}$, and $\frac{t}{\tau_{\rm open}}$.

A crucial issue is, of course, to determine the ranges of values
of the parameters (opening time, oscillation frequencies and thermalization time) relevant to our system. This is however not so straightforward, since these parameters vary over considerably large ranges \cite{Frauenfelder91}. They depend on the type of molecule we consider, on its size, on the size of the active site, on which parts of the active site contract (i.e. the whole site or only subparts involving some of the functional groups), and on the degree of interaction with the environment. Furthermore, such parameters are typically not all available for the same molecule, but some parameters are known for some molecules, other parameters for other molecules.

For the figure, we have chosen the following values:
$\frac{\hbar\omega_0}{k_{B}T}=0.032$, $\frac{\omega_1}{\omega_0}=2$, $\gamma\tau_{\rm open}=1$.
At room temperature, $T=300K$, this corresponds to molecular oscillation times $\tau_{\rm osc} = 2\pi/\omega_{0}=5$ps, $\tau_{\rm osc}' = 2\pi/\omega_{1}=2.5$ps, which is a typical time scale of elastic vibrations. 
Possible values for the opening time $\tau_{\rm open}$, which is the time for (part of) a conformational transition, may range from fractions of nanoseconds 
up to microseconds and beyond, depending on the size of the protein and the specific type of transition. Similarly, the thermalization time $\tau_{\rm therm}\equiv\gamma^{-1}$ may vary significantly, depending on coupling of the oscillators to the environment. For fast configurational processes, an exemplary choice would be $\tau_{\rm open}=100$ps as the time scale for re-opening the cavity and $\gamma^{-1}=100$ps as the thermalization time for the vibrations.

As we have mentioned, these values will depend on the specific molecular realization, and all of these numbers are subject to considerable variation \cite{RecentReviews}. But it seems reasonable to assume that the time scales of the conformational transition and of the cavity vibrations are separated by at least one order of magnitude, i.e. $\tau_{\rm osc}, \tau_{\rm osc}' \ll \tau_{\rm open}$ (which is the so-called adiabatic regime, with $\dot\omega/\omega \ll 1$). The timescale for $\tau_{\rm open}$ may also be much larger; likewise, the timescale for thermalization may be different. Whether or not the refrigeration effect can be observed depends on the timescale of the thermalization relative to the opening time, i.e. on $\tau_{\rm open}/\tau_{\rm therm}=\gamma\tau_{\rm open}$. As long as the thermalization time is larger or at least comparable to the opening time, i.e. $\gamma\tau_{\rm open}\lesssim 1$, the effect can clearly be observed, as demonstrated in Fig.~\ref{RefrigerationDynamicsC}.
One can seen how the effective temperature of the cavity $T(t)$ drops during the opening transition to about $65\%$ of the environmental temperature $T$. After the cavity has reached the open position, its temperature relaxes to the environmental temperature. It should however be noted that a useful effect of refrigeration may persist up to several multiples of $\tau_{\rm open}$: A refrigeration by about 1 degree Kelvin, for example, means a relative value $T(t)/T=299/300=0.997$ which is reached only after about $6\times\tau_{\rm open}$. Of course, the effect is more persistent for longer thermalization times.

\vspace*{0.5cm}

\vspace*{3pt}{\bf Conclusions}. The mechanism for intra-molecular refrigeration that we have discussed in this paper is simple. It is based on conformational changes that induce time-dependent forces between different parts of a molecule, reducing their relative vibrations. It works as long as the configurational changes, which may involve only small parts of a protein, happen faster or at least on a comparable timescale as the damping of the relevant vibrations. The changes in local temperature, which can be generated by this mechanism, are, even for moderate variations of the binding potential, quite significant: It should be remembered that, for many biological processes, a refrigeration by only a few degree Kelvin can make all the difference.

Many variations of this scheme are conceivable. For example, it is possible that cooling due to conformational variations may take place at other sites, near the active site, but without changing the latter's shape. Through heat transfer to the cooling centers, effective refrigeration of the active site would be possible without interrupting the catalyzer process.

The improvement of enzyme functionality is but one example
of the potential applications of intra-molecular refrigeration, but there should be many others. Molecular refrigeration might also play a role e.g.\ in molecular machines, to stabilize their operation against the accumulation of noise and errors, or in more complex processes such as those involved in protein synthesis.

Last, but not least, the possibility of maintaining temperature gradients on a molecular scale is by itself of considerable interest for the investigation of thermodynamical processes at the bio-molecular level.


\begin{thebibliography}{99}
\bibitem{RecentReviews}
For recent reviews, where one can find a discussion of some of the timescales, see e.g. \cite{Agarwal06,McKenzie08} and references cited therein.
\end{thebibliography}
\end{document}